# Counterintuitive VSM Behavior under CVR Incorporating Distribution System

Alok Kumar Bharati, *Student Member, IEEE*, Venkataramana Ajjarapu, *Fellow, IEEE,* Zhaoyu Wang *Member, IEEE*

***Abstract*—This paper analyses the impact of conservation by voltage reduction (CVR) on voltage stability margin (VSM) considering transmission and distribution (T&D) systems. VSM is determined by P-V curve analysis using PSSE and GridLAB-D solvers to co-simulate the T&D systems under CVR and No CVR conditions. ZIP loads with profile [ZIP] = [0.4 0.3 0.3] are used to model the load. The paper discusses the counterintuitive result: under CVR, the VSM is reduced. Theoretical justification for the reduced VSM under CVR is the increase in the effective impedance between generation and load and this is proved using an extended 2-bus system. The paper shares T&D co-simulation results with IEEE 9-bus transmission system and a larger 123-bus distribution system and with distributed generation (DG) in unity power factor (UPF) and volt-VAR control (VVC) mode.**

## I. Introduction

POWER system is generally divided into two sub-systems i.e., the transmission network which is a complex meshed network, operating at higher voltages and the distribution network which is usually radial in nature, operating at lower voltages where loads are connected. In the real world, the loads are not constant and vary throughout the day and vary in their profiles throughout the year. Peak loads have higher magnitude of power and last for a relatively shorter duration. Addressing the peak loads is technically and economically challenging for the power system operators. The cost of meeting the peak loads is usually transferred to distribution utilities in the form of peak pricing. The utilities have several programs aimed at reducing the peak loads to reduce the power purchased at peak pricing.

A significant part of the load is voltage dependent load [1] and is modeled as ZIP loads. Load with a ZIP profile, [Z I P] = [0.4 0.3 0.3] means 40% is constant impedance (Z) load, 30% is constant current (I) and 30% is constant power (P) loads.

$$P_{ZIP} = P_0\left(P_Z\left(\frac{V}{V_0}\right)^2 + P_I\left(\frac{V}{V_0}\right) + P_P\right) \quad (1)$$

$$Q_{ZIP} = Q_0\left(Q_Z\left(\frac{V}{V_0}\right)^2 + Q_I\left(\frac{V}{V_0}\right) + Q_P\right) \quad (2)$$

Where,

$V$ →Operating voltage;

$V_0$ →Nominal voltage;

$P_0, Q_0$ →Base real and reactive powers (P & Q) of that load;

$P_Z, Q_Z$ →Constant impedance fractions of P & Q;

$P_I, Q_I$ →Constant current fractions of P & Q;

$P_P, Q_P$ →Constant power fractions of P & Q;

$P_Z + P_I + P_P = 1; Q_Z + Q_I + Q_P = 1;$

Equations (1) and (2) define the ZIP load model. The load can be reduced by reducing voltage on distribution feeders thereby reducing the voltage dependent loads. This method of reducing load is popularly termed as conservation by voltage reduction (CVR) and is mostly used to reduce the peak load in a feeder [2]. However, the voltage can only be reduced to be within the ANSI standard voltage limit [3]. CVR has been successfully deployed by many utilities and has been studied by academia [2].

It is important to understand that the utility deploys CVR very conservatively only during peak pricing and so, CVR can also be expanded as conservative voltage reduction. CVR is essentially a distribution system phenomenon and needs to consider detailed distribution system models and solvers for analysis [4]. There have been several studies conducted to study CVR and its benefits [5]-[8], however, these are based on real power impact. CVR also affects reactive power and its impact has not been studied in detail.

### A. Motivation

By intuition, there are two important aspects of CVR that are closely related to this study on CVR impact on VSM:

1. Load reduction under CVR should allow for further load increase and help to increase the VSM.
2. The reactive demand reduction increases the voltage, which may also help the VSM.

These intuitions motivate the present investigation of impact of CVR on voltage stability margin (VSM) considering the transmission and distribution (T&D) systems together.

### B. CVR and Long-Term Voltage Stability

There is one paper that reports that CVR improves the voltage stability of the system [9], but it does not consider any distribution system model in this study. Our previous work published in reference [10] on CVR impact on VSM with aggregated and de-aggregated load shows that for the

This paper is submitted for review to IEEE Power Engineering Letters on 12/23/2019. This work is funded by Power Systems Engineering Research Center (PSERC) project S-70

Alok Kumar Bharati is a PhD student at the Department of Electrical and Computer Engineering, Iowa State University, Ames, IA 50010 USA (e-mail: alok@ iastate.edu).

Dr. Venkataramana Ajjarapu is David C Nicholas endowed Professor with the Department of Electrical and Computer Engineering, Iowa State University, Ames, IA 50010 USA (e-mail: vajjarap@ iastate.edu).

Dr. Zhaoyu Wang is Harpole-Pentair Assistant Professor with the Department of Electrical and Computer Engineering, Iowa State University, Ames, IA 50010 USA (e-mail: wzy@ iastate.edu).

aggregated load case, i.e., only with transmission system analysis, CVR increases load margin by 1.835% (similar to [9]) whereas, for the same with de-aggregated load, the increase in margin due to CVR is reduced to 0.399%. Reference [10] suggests using T&D co-simulation to accurately understand the true impact of CVR on VSM.

There is little done to understand the impact of CVR on reactive power and in turn, voltage stability margin of the system considering detailed transmission and distribution (T&D) system models. Refernce [11] shows that T&D Co-Simulation is an effective tool to represent transmisison and distributoin systems in detail for VSM assessment.

## II. CVR Impact on VSM Through T&D Co-Simulation

To include transmission and distribution system models together, we use T&D co-simulation method. The transmission systems and distribution systems have different characteristics that need different solvers to analyze the transmission and distribution systems.

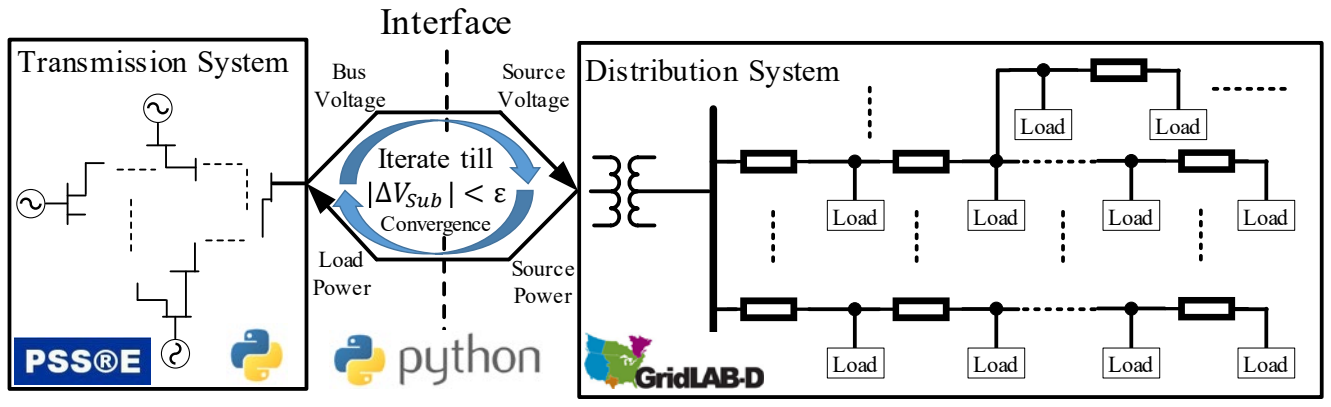

Fig. 1. T&D Co-Simulation Framework to Study CVR Impact on VSM.

A simple interface is developed in Python to efficiently co-simulate commercial solvers that analyse transmission and distribution systems. The Python interface is programmed to exchange the data between the transmission and distribution system solvers and ensure convergence of every operating point co-simulated as described in [11]. PSSE and GridLAB-D solvers are used for the transmission and distribution system respectively and are co-simulated as shown in Fig. 1.

### A. *CVR Impact on VSM using T&D Co-Simulation*

To assess the CVR impact more accurately, T&D co-simulation is condsidered. Reference [11] also advises to co-simulate T&D system together to capture the details of the distribution system accurately for VSM studies. Fig. 2 shows the P-V curves for the IEEE 9-bus transmission (9-T) system with IEEE 4-bus distribution (4-D) system at load bus for CVR and No CVR cases. The load on the system is modeled as a ZIP load with ZIP profile [ZIP]=[0.4 0.3 0.3]. CVR is simulated by changing taps on the substation transfomer secondary.

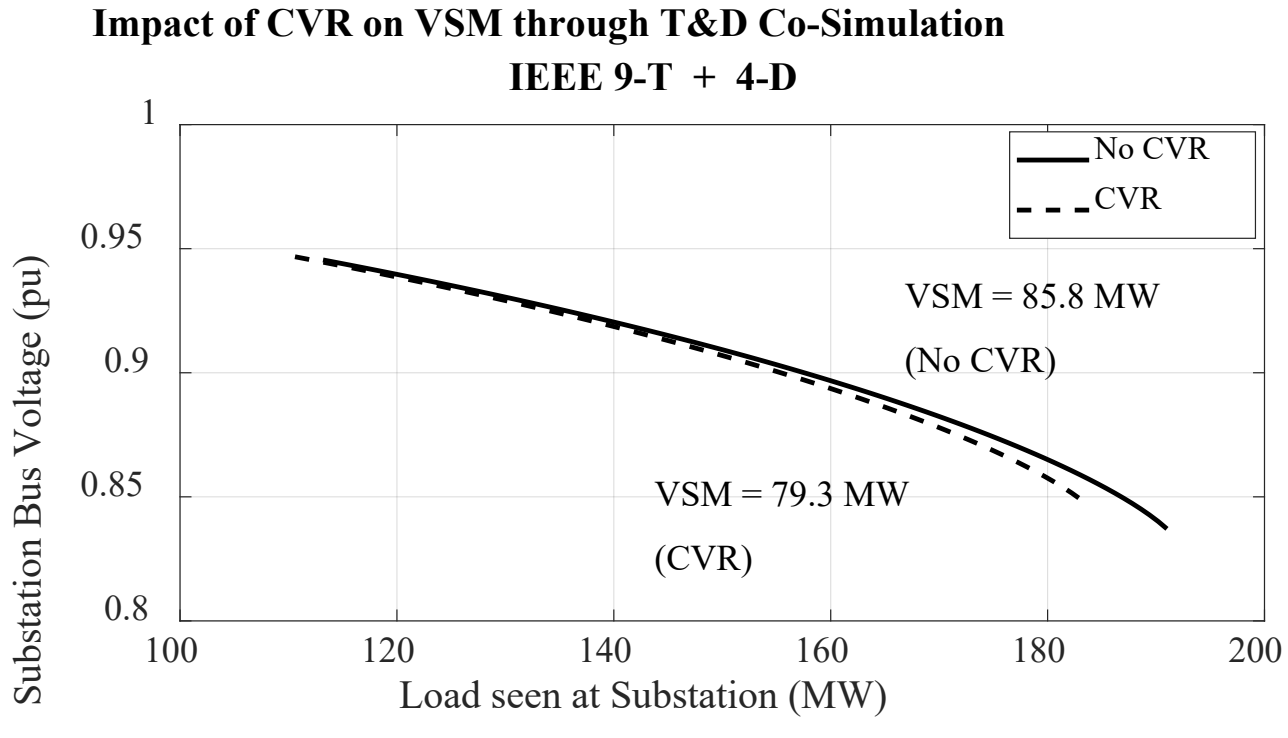

Fig. 2. PV- Curves for IEEE 9-T and IEEE 4-D for No CVR and CVR

From Fig. 2, we can see for the case with CVR, due to the reduction in voltage, the load is also reduced, and transmission voltage is higher compared to the No CVR case. However, the VSM is reduced by 7.5% which is counterintuitive.

### B. *Results with Larger 123-Bus Distribution System with DG*

The IEEE 123-Bus distribution system is considered to be co-simulated with the IEEE 9-Bus transmission system to study the impact of CVR on VSM of the larger system. 60% DG in the form of solar photovoltaic (PV) is considered on the IEEE 123-bus distribution system. DG is operating in unity power factor (UPF) mode and in volt-var control (VVC). We perform CVR in a similar way as described earlier, but, in the case of DG in VVC mode, the voltage setpoint of inverters is set to 1 pu as compared to 1.05 pu for No CVR case. The IEEE 123-Bus system is modified by converting the 3-phase loads to three 1-phase loads to add the DG and PV inverters. The VVC implemented is the standard droop control as per IEEE 1547.

TABLE I shows the T&D co-simulation results of CVR impact on VSM with IEEE 9-T and 123-Bus distribution system (123-D) for cases with no DG, DG in UPF and DG in VVC.

TABLE I. Influence of DG on Impact of CVR on VSM with IEEE 9-T + 32 x modified 123-D Co-Simulation

| No CVR/ CVR | DG (mode) | VSM (MW) | % Reduction in VSM under CVR |
|---|---|---|---|
| No CVR | No DG | 226.1306 | 7.39 % |
| CVR | No DG | 209.4102 | |
| No CVR | 60% DG (UPF) | 268.2128 | 5.37 % |
| CVR | 60% DG (UPF) | 253.8141 | |
| No CVR | 60% DG (VVC) | 304.3152 | 11.67 % |
| CVR | 60% DG (VVC) | 268.785 | |

## III. Analysis of Counterintuitive VSM Behavior under CVR Deployment

To analyze the CVR impact on VSM, let us consider a simple 2-bus system i.e., load connected to an infinite bus (Bus1 in Fig 3) through a line, but, with a distribution system transformer to simulate CVR as shown in Fig. 3 (a), (b). System in Fig. 3(a) does not have any distribution system but, Fig. 3(b) shows system with an equivalent distribution feeder representing distribution system.

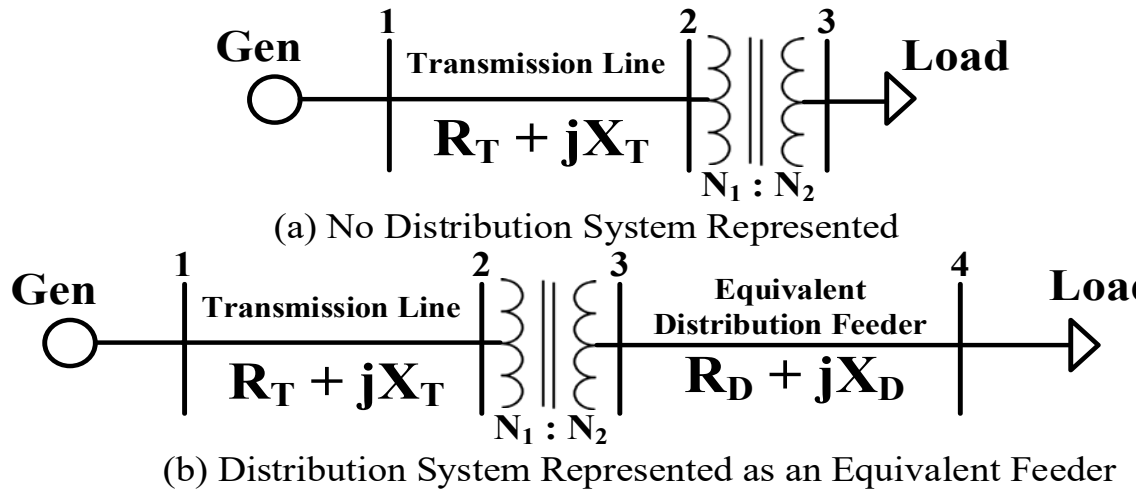

Fig. 3. A Simple 2-Bus System with and without Distribution Equivalent Distribution Feeder

### A. *Importance of Distribution System Representation*

We will use the systems shown in Fig. 3(a), 3(b) to study the CVR impact on VSM of the system using continuation power flow (CPF) described in [12] to show the importance of modeling distribution system for this study. For the simple system the the transmission line impedance is considered as $R_T + j\, X_T = 0.01 + j\, 0.06\, pu$ . The load is modeled as 60MW and 20MVAR using the ZIP profile [ZIP]=[0.4 0.3 0.3]. The load is increased using a parameter $'\lambda'$ to draw the $\lambda - V$ curve to determine the VSM of the system using CPF. Fig. 4 shows the results for the system shown in Fig. 3 (a). The secondary taps are reduced from 1 pu to 0.95 pu to simulate CVR. It can be seen from Fig. 4 that under CVR, the VSM of the system is higher accoring to the intuition, but, there is no distribution system representation. Let us consider adding distribution system.

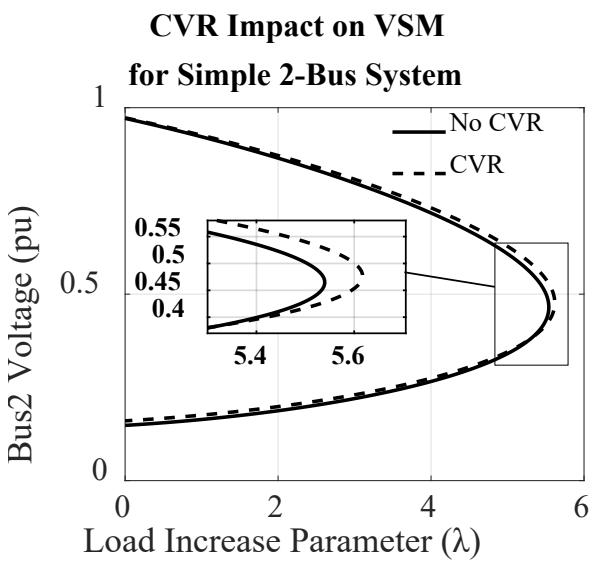

Fig. 4 Voltage Stability Curve for 2-bus system (Fig. 3 (a)) (Intuitive)

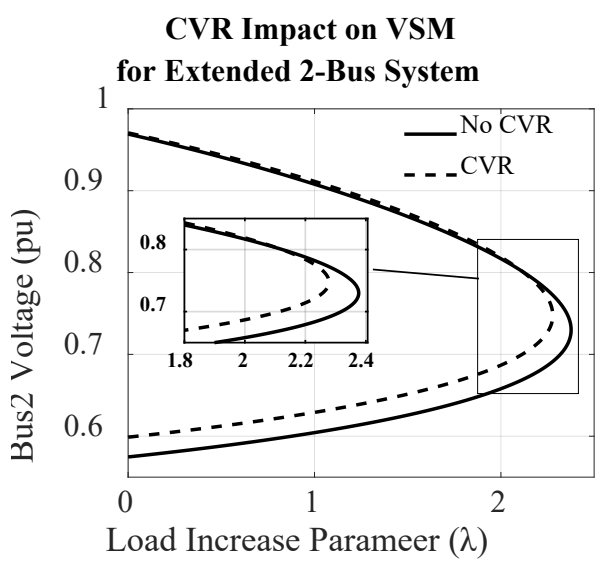

Fig. 5. Voltage Stability Curve for Extended 2-bus system (Fig. 3 (b)) (Counterintuitive)

The extended 2-bus system with the eq. D-Feeder is shown in Fig. 3 (b). The parameters of the eq. D-Feeder are computed for the IEEE 4-bus distribution system by computing the losses with the substation voltage at 1-pu. The $R_D + jX_D = 0.03 + j0.06\, pu$ is computed using method in reference [11]. The CVR impact on VSM result for the extended 2-bus system with the eq. D-Feeder is shown in Fig. 5. It shows under CVR, the VSM of the system is reduced which is counterintuitive.

### *B. Thoeritical Understanding of Counterintuituve VSM behavior under CVR*

We use the distribution substation transformer to simulate CVR by reducing the turns of the secondary winding from $\mathrm{N}_2$ to $\mathrm{N}_2'$. Superscript $'$ denotes the physical quantity under CVR. Let the turns ratio of the transformer be 'k', where:

$$k = \frac{N_1}{N_2} \quad (3)$$

The equivalent impedance between the generator bus, 'Bus 1' to the load bus, 'Bus 4' is given by:

$$Z_{eq} = (R_T + jX_T) + k^2(R_D + jX_D) \quad (4)$$

Now, CVR deployment is simulated by reducing the number of secondary windings of the distribution transformer:

$$k' = \frac{N_1}{N_2'} \quad (5)$$

$$N_2' < N_2 => \ k' > k \quad (6)$$

$$Z_{eq}' = (R_T + jX_T) + k'^2(R_D + jX_D) \quad (7)$$

From equations (4) and (7), $Z_{eq}'$ has higher resistance and reactiance than $Z_{eq}$. Therefore, the effective impedance between generator and load is higher under CVR deployment. The increased impedance directly affects the VSM of the system. This can only be captured by representing the distribution system along with the tramsission system. If $R_D + jX_D$ is zero, the increase is impedance is not reflected and only load decrease is captured like in Fig. 4. But Fig. 5 shows the real impact and is counterintuitive.

## IV. Conclusions

This paper highlights and re-emphasizes the need to consider the distribution network in detail for performing accurate system studies involving distribution system control like CVR. The results are counterintuitive and show that: VSM reduces under CVR deployment, however, the transmission system voltage increases due to the reduction in load. The paper does not discourage CVR in anyway but, provides awareness of the true impact of CVR on VSM. This behavior can be captured accurately through T&D co-simulation.

## Acknowledgment

This work is funded by Power Systems Engineering Research Centre (PSERC) project S-70. The authors also thank the contributors of Matpower and GridLAB-D software.